\documentclass{iau}
\usepackage{graphicx}

\title[The Origin of GC Diffuse X-ray Emission] 
{The Origin of the Galactic Center Diffuse X-ray Emission 
Investigated by Near-infrared Imaging and Polarimetric Observations}

\author[Nishiyama et al.]   
{Shogo Nishiyama$^1$, 
Kazuki Yasui$^2$, Tetsuya Nagata$^2$, Tatsuhito Yoshikawa$^2$, 
Hideki Uchiyama$^3$, \and Motohide Tamura$^1$}

\affiliation{$^1$National Astronomical Observatory of Japan, 
Mitaka, Tokyo 181-8588, Japan
\\ email: {\tt shogo.nishiyama@nao.ac.jp} \\[\affilskip]
$^2$Department of Astronomy, Kyoto University, 
Kyoto 606-8502, Japan\\
$^3$Science Education, Faculty of Education, Shizuoka University, 
Shizuoka 422-8529, Japan
}

\pubyear{2013}
\volume{303}  
\pagerange{???--???}
\jname{The Galactic Center: Feeding and Feedback in a Normal Galactic Nucleus}
\editors{}
\begin{document}

\maketitle

\begin{abstract}
The origin of the Galactic center diffuse X-ray emission (GCDX)
is still under intense investigation.
We have found a clear excess in a longitudinal GCDX profile
over a stellar number density profile in the nuclear bulge region,
suggesting a significant contribution of diffuse, interstellar hot plasma to the GCDX.
We have estimated that
contributions of an old stellar population to the GCDX
are $\sim 50$\,\% and $\sim 20$\,\% 
in the nuclear stellar disk and nuclear star cluster, respectively.
Our near-infrared polarimetric observations show that
the GCDX region is permeated by a large scale, toroidal magnetic field.
Together with observed magnetic field strengths in nearly energy equipartition,
the interstellar hot plasma could be confined by the toroidal magnetic field.
\end{abstract}

\firstsection 
\section{Introduction}

Past X-ray observations have revealed the presence of 
a diffuse 6.7\,keV emission from highly ionized, He-like Fe ions
toward the Galactic center
(\cite[e.g., Koyama et al. 1989]{89Koyama}).
The line emission and an associated continuum component,
called the Galactic center diffuse X-ray emission (GCDX),
resembles the Galactic ridge diffuse X-ray emission (GRXE).
For the GRXE,
more than $80\,\%$ of the diffuse emission 
has been claimed to be resolved
into point sources (\cite[Revnivtsev et al. 2009]{09Revnivtsev}),
suggesting faint X-ray point sources in origin.

However, the origin of the GCDX, in particular its very hot component 
with a temperature of $kT \sim 7$\,keV, is more puzzling.
Only 10 - 40\,\% of the GCDX has been claimed to be resolved
into faint point sources so far (\cite[e.g., Muno et al. 2004]{04Muno}).
The plasma temperature of the GCDX
is likely to be systematically higher than the GRXE
(\cite[Yamauchi et al. 2009]{09Yamauchi}).
These results indicate a different origin of the GCDX from the GRXE.

Another idea for the origin of the GCDX is
a diffuse, interstellar hot plasma
(\cite[e.g., Koyama et al. 1989]{89Koyama}).
In this case, the spatial distribution of the GCDX
is expected to be different from that of stellar populations.
Hence we have constructed a stellar number density map
of the Galactic center region from new near-infrared (NIR) imaging observations. 
Here we summarize results of the NIR observations,
and provide additional evidence for the hypothesis
that the GCDX arises from a truly diffuse hot plasma.

We also show results of our NIR polarimetric survey
to reveal magnetic field configuration in the Galactic center region.
The results provide strong evidence for a large-scale
toroidal magnetic field configuration in the GCDX region,
implying a magnetic field confinement of the interstellar hot plasma.

\section{Stellar Number Density Profile}

The central region of our Galaxy,
$\mid l \mid \lesssim 3 .\!\!^{\circ} 0$ and $\mid b \mid \lesssim 1 .\!\!^{\circ} 0$
was observed from 2002 to 2004
using the NIR camera SIRIUS on the 1.4 m telescope IRSF.
After removing foreground/background sources,
we carry out an extinction correction for detected stars 
(\cite[Nishiyama et al. 2006]{06Nishiyama}),
and obtain an extinction-corrected $K_S$-band magnitude, $K_{S,0}$ for each star.
Taking into account completeness limits,
we construct a stellar number density map
using stars with $K_{S,0} <10.5$.
For more detail, see \cite[Nishiyama et al. (2013)]{13Nishiyama}.

The profile of the 6.7\,keV line emission along the Galactic plane
measured by {\it Suzaku} 
(\cite[Koyama et al. 2007, Uchiyama et al. 2011]{07Koyama,11Uchiyama})
clearly show an excess at the nuclear bulge (NB) region 
($\mid l \mid \lesssim 1 .\!\!^{\circ} 0$)
over the stellar number density profile 
(Fig. 1, top panel).
The two profiles are over-plotted and scaled
to have the same values at 
$1 .\!\!^{\circ} 5 < \mid l_* \mid < 2 .\!\!^{\circ} 8$
($l_*$ denotes the angular distance from Sgr\,A* 
along the Galactic longitude).

\begin{figure}[h]
\begin{center}
 \includegraphics[width=3.5in]{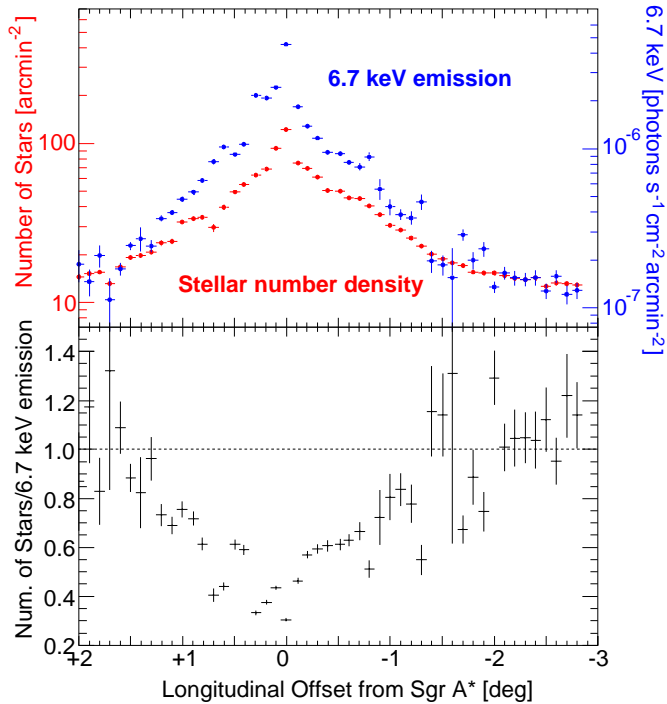} 
 \caption{
Top: Longitudinal profiles of the stellar number density (red)
and the 6.7\,keV emission
(\cite[blue; Koyama et al. 2007;]{07KoyamaFe} 
\cite[Uchiyama et al. 2011]{11Uchiyama}).
The region in the Galactic bulge, 
$1 .\!\!^{\circ} 5 < \mid l_* \mid  < 2 .\!\!^{\circ} 8 $,
is used to scale the the 6.7\,keV emission profile
to have the same value as the stellar number density.
Bottom: Longitudinal profile
of the ratio of the stellar number density to the 6.7\,keV emission,
scaled to be unity at the position for normalization.
This profile represents a contribution of point sources,
traced by our NIR observations, 
to the GCDX in the assumption that the contribution of 
truly diffuse hot plasma is negligible 
at the position for normalization (i.e., the Galactic ridge region).
}
   \label{fig1}
\end{center}
\end{figure}

The majority of faint X-ray 
($L_{\mathrm{2-10\,keV}} < 10^{30}$\,erg\,s$^{-1}$) sources 
which have not been resolved
but are expected to contribute to the GCDX are
old binary systems such as cataclysmic variables.
To investigate the contribution of the point sources,
especially of the old stellar population detectable in our observations,
we construct a longitudinal profile for the ratio
of  the stellar number density to the 6.7\,keV emission
(Fig. 1, bottom panel).
When the profile is scaled to be unity at the positions of normalization, 
$1 .\!\!^{\circ} 5 < \mid l_* \mid < 2 .\!\!^{\circ} 8$,
the ratios are $\sim 0.6 - 0.7$ and $\sim 0.3$ 
in the nuclear stellar disk (NSD) and the nuclear star cluster (NSC), respectively.
Considering different star formation histories 
in the NSD, NSC, and  the Galactic Bulge (GB),
the contributions of the old stellar population to the GCDX are estimated to be
$\sim 0.5$ and $\sim 0.2$ for the NSD and NSC, respectively,
in the assumption that
the contribution of point sources
in the GB is 100\,\%.

The difference of the longitudinal profiles (Fig. 1, top panel)
is difficult to be explained by a variation of an emissivity of the point sources.
In \cite[Nishiyama et al. (2013)]{13Nishiyama}, we show that 
different initial mass functions, binary fractions, and star formation histories
in the NSD and NSC from the GB
cannot explain the difference in the longitudinal profiles.

Another factor which can change the emissivity is a Fe abundance.
A higher Fe abundance of stars in the NB region than the GB 
leads to a higher X-ray emissivity.
However, the 6.7\,kev emission is 
$\sim 1.6$ times larger than the stellar number density in the NSD
when they are shifted to have the same value at $\mid l_* \mid \sim 2^{\circ} $.
It means that the required gradient of the Fe abundance is
$\sim 60\% / 2^{\circ}  \sim 20 \% / 100$\,pc.
This gradient is much larger than that in the inner Galaxy;
an observed gradient of the Fe abundance at the inner Galaxy
is $\sim$ 0.06\,dex/kpc, which corresponds to only a few \%/100\,pc
(\cite[Gonz{\'a}lez-Fern{\'a}ndez et al. 2008]{08Gonzalez}).
In addition, the stellar Fe abundance in the Galactic center is not so high.
Recent observations show a nearly solar Fe abundance even for young stars
(\cite[e.g., Davies et al. 2009, Najarro et al. 2009]{09Davies,09Najarro}).
Thus the abundance for old stars which could be the origin of the GCDX
is expected to be lower than the solar Fe abundance.

\section{Magnetic Field Configuration}

The most puzzling aspect of the GCDX is its high temperature.
Since the $kT \approx 7\,$keV plasma is too hot to be gravitationally bound,
it requires a huge energy source without the confinement of the plasma.
One idea to address this energetic issue is a confinement
by magnetic fields (\cite[MFs; Makishima 1994]{94Makishima}).
If a large-scale toroidal MF is developed and sustained, 
and the MF is strong enough for nearly energy equipartition 
with the plasma of $kT \sim 7$\,keV,
the GCDX could be almost confined within the NB.

To determine the large-scale interstellar MF configuration, 
we have carried out NIR polarimetric observations for 
$\mid l \mid \lesssim 1 .\!\!^{\circ} 5$ and $\mid b \mid \lesssim 1 .\!\!^{\circ} 0$
using IRSF and 
a NIR polarimetric imager SIRPOL  from 2006 to 2010.
Comparing the polarization between stars distributed
further and closer side in the Galactic center,
we obtain polarization originating from
magnetically aligned dust grains in the Galactic center
(\cite[for more detail, see Nishiyama et al. 2009, 2010]{09Nishiyama,10Nishiyama}).
The polarized angle traces the MF direction in the Galactic center 
projected onto the sky (Fig. 2).

The obtained polarization map (Fig. 2) suggests 
a large-scale toroidal MF configuration in the NB. 
The histogram of the MF directions at $\mid b \mid < 0 .\!\!^{\circ} 4$ 
has a clear peak at $90^{\circ} $ 
(\cite[see Fig. 5 in Nishiyama et al. 2013]{13Nishiyama})
which is the direction parallel to the Galactic plane.
At high Galactic latitude 
($\mid b \mid \gtrsim 0 .\!\!^{\circ} 4$),
the fields are nearly perpendicular to the plane,
i.e., poloidal configuration.
This suggests a transition of the large-scale configuration,
and such a transition can be naturally explained by
the time evolution of MFs.
The transition region, 
$\mid b \mid \sim 0 .\!\!^{\circ} 3 - 0 .\!\!^{\circ} 4$, is in good agreement 
with the scale height of the 6.7\,keV emission, 
$0 .\!\!^{\circ} 27$ (\cite[Uchiyama et al. 2013]{Uchiyama13}).

If the plasma were not confined, 
it would be rushing out of the Galactic plane vertically as a galactic wind.
The escape velocity, typically several hundred km\,s$^{-1}$,
is smaller than the sound speed of the $7\,$keV hot plasma
of $> 1000$\,km\,s$^{-1}$.
Assuming the gas flows out from the X-ray emitting region 
at the sound speed,
the escape timescale is $\sim 4 \times 10^4$\,yr 
(\cite[Belmont et al. 2005]{05Belmont}),
requiring a huge energy input to sustain the hot plasma.
If the plasma is confined magnetically, 
and there is no other cooling mechanism, 
the hot plasma only cools by radiation with a timescale of 
$10^7 - 10^8$\,yr (\cite[Muno et al. 2004]{04Muno}),
several orders of magnitude longer than the escape timescale.
This would reduce the required energy input and thus relax the energetic problem.

\begin{figure}[b]
\begin{center}
 \includegraphics[width=4.2in]{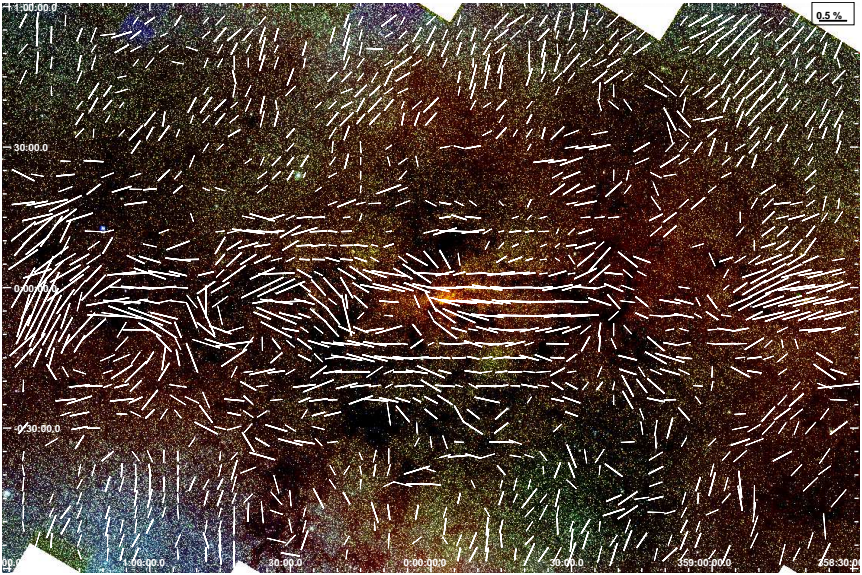} 
 \caption{
Near-infrared mosaic image of the Galactic center region
covering $3 .\!\!^{\circ} 0 \times 2 .\!\!^{\circ} 0$ in the Galactic coordinate,
taken with the IRSF telescope and NIR camera SIRIUS.
The three NIR bands are $J$ (blue, $1.25\mu$m), 
$H$ (green, $1.63\mu$m), and $K_S$ (red, $2.14\mu$m).
Observed polarization angles for the Galactic center components 
in the $K_S$ band are also plotted 
with white bars whose length indicates the degree of polarization.
}
   \label{fig2}
\end{center}
\end{figure}

\end{document}